\documentstyle[12pt]{article}
\catcode`\@=11
\font\teneus=eusm10 scaled \magstep1
\font\seveneus=eusm7 scaled \magstep1
\font\fiveeus=eusm5 scaled \magstep1

\newfam\eusfam
\textfont\eusfam=\teneus  \scriptfont\eusfam=\seveneus
  \scriptscriptfont\eusfam=\fiveeus

\def\hexnumber@#1{\ifnum#1<10 \number#1\else
 \ifnum#1=10 A\else\ifnum#1=11 B\else\ifnum#1=12 C\else
 \ifnum#1=13 D\else\ifnum#1=14 E\else\ifnum#1=15 F\fi\fi\fi\fi\fi\fi\fi}

\def\Cl{\ifmmode\let\next\Cl@\else
 \def\next{\errmessage{Use \string\Cl\space only in math mode}}\fi\next}
\def\Cl@#1{{\Cl@@{#1}}}
\def\Cl@@#1{\fam\eusfam#1}

\catcode`\@=\active

\catcode`\@=11

\font\teneuf=eufm10 scaled \magstep1
\font\seveneuf=eufm7 scaled \magstep1
\font\fiveeuf=eufm5 scaled \magstep1

\newfam\euffam
\textfont\euffam=\teneuf  \scriptfont\euffam=\seveneuf
  \scriptscriptfont\euffam=\fiveeuf

\def\hexnumber@#1{\ifnum#1<10 \number#1\else
 \ifnum#1=10 A\else\ifnum#1=11 B\else\ifnum#1=12 C\else
 \ifnum#1=13 D\else\ifnum#1=14 E\else\ifnum#1=15 F\fi\fi\fi\fi\fi\fi\fi}

\def\Got{\ifmmode\let\next\Got@\else
 \def\next{\errmessage{Use \string\Got\space only in math mode}}\fi\next}
\def\Got@#1{{\Got@@{#1}}}
\def\Got@@#1{\fam\euffam#1}

\catcode`\@=\active

\catcode`\@=11

\font\tenmsx=msxm10 scaled \magstep1
\font\sevenmsx=msxm7 scaled \magstep1
\font\fivemsx=msxm5 scaled \magstep1
\font\tenmsy=msym10 scaled \magstep1
\font\sevenmsy=msym7 scaled \magstep1
\font\fivemsy=msym5 scaled \magstep1
\newfam\msxfam
\newfam\msyfam
\textfont\msxfam=\tenmsx  \scriptfont\msxfam=\sevenmsx
  \scriptscriptfont\msxfam=\fivemsx
\textfont\msyfam=\tenmsy  \scriptfont\msyfam=\sevenmsy
  \scriptscriptfont\msyfam=\fivemsy
\def\hexnumber@#1{\ifnum#1<10 \number#1\else
 \ifnum#1=10 A\else\ifnum#1=11 B\else\ifnum#1=12 C\else
 \ifnum#1=13 D\else\ifnum#1=14 E\else\ifnum#1=15 F\fi\fi\fi\fi\fi\fi\fi}

\def\Bbb{\ifmmode\let\next\Bbb@\else
 \def\next{\errmessage{Use \string\Bbb\space only in math mode}}\fi\next}
\def\Bbb@#1{{\Bbb@@{#1}}}
\def\Bbb@@#1{\fam\msyfam#1}

\catcode`\@=\active

\begin{document}

\newcommand{\ie}{{\em i.e.}}
\newcommand{\qslc}{U_q {\Got sl}(2,{\Bbb C})}
\newcommand{\slc}{{\Got sl}(2,{\Bbb C})}
\newcommand{\SLC}{SL(2,{\Bbb C})}
\newcommand{\SLR}{SL(2,{\Bbb R})}
\newcommand{\be}{\begin{equation}}
\newcommand{\ee}{\end{equation}}
\newcommand{\half}{\frac{1}{2}}
\newcommand{\zb}{\bar{z}}
\newcommand{\bemath}{\begin{displaymath}}
\newcommand{\eemath}{\end{displaymath}}
\newcommand{\cJ}{{\cal J}}
\newcommand{\qvalue}{\frac{p'}{p}}
\newcommand{\vp}{\varphi}
\newcommand{\cp}{\varphi_{cl}}
\newcommand{\DD}{{\Got D}}


\rightline{July, 1994}
\vspace{1.5cm}
\begin{center}
\LARGE{Towards a Strong Coupling Liouville Gravity}\\[2em]
\large{Takashi Suzuki}\footnote{Present address,
YITP-Uji, University of Kyoto, Uji 611, Japan}  \\[1em]
\normalsize     Department of Mathematics and Statistics \\
\normalsize              University of Edinburgh         \\
\normalsize                  Edinburgh,  U.K.
\end{center}
\pagestyle{plain}

\vspace{0.5cm}

\begin{abstract}
A possibility of strong coupling quantum Liouville gravity is investigated via
infinite dimensional representations of $\qslc$ with $q$ at a root of unity.
It is explicitly shown that vertex operator in this model can be written
by a tensor product of a vertex operator of the classical Liouville theory
and that of weak coupling quantum Liouville theory.
Some discussions about the strong coupling Liouville gravity within this
formulation are given.
\end{abstract}

\vspace {1.0cm}

In the recent developments of $2D$ quantum Liouville gravity theory\cite{Se}
as a non-critical string theory\cite{Po1}-\cite{DK},
one of the important features is that
it has quantum group structure\cite{Gerv,ST}.
Precisely, the vertex operators in the theory can be expressed
in terms of the representations of  $\qslc$.
Upon remembering that there are two kinds of representations
of $\qslc$, namely, of finite dimension and of infinite one,
and that they are completely different from each other,
we can expect that there are two phases in the Liouville gravity.
These phases correspond to, so-called, weak coupling regime and
strong coupling regime.
A number of works have revealed remarkable features of the weak coupling
Liouville gravity where finite dimensional representations appear.
However, in the weak coupling regime,
the well-known restriction, that is, the $D=1$ barrier has bothered us.

On the other hand, a successful work for the strong coupling Liouville gravity
has been done by Gervais and his collaborators\cite{Ge}.
In these works, they made use of the infinite dimensional representations
of $\qslc$ with {\rm generic} $q$ and have
shown that consistent theories can be built provided
the central charge of the Liouville gravity is 7, 13 or 19.
However, up to now, no work has been done for the quantum Liouville gravity
based on the infinite dimensional representations of $\qslc$ with $q$
at a root of unity.
Hereafter we will denote by $\DD$ the infinite dimensional representations
of $\qslc$ when $q$ is a root of unity.
We can expect that the theory with such a $q$ will be drastically
different from that with a generic $q$.
Indeed, in \cite{MS} it has been shown that $\DD$ is completely different
from representations with generic $q$.
It is, therefore, quite interesting and worthwhile to investigate
the structure of the quantum Liouville theory via such representations.
The importance of the Liouville gravity based on $\DD$ is announced
also in Ref.\cite{ST}.
In this letter we will examine the structure of the strong coupling
Liouville gravity when we apply $\DD$ instead of the finite dimensional
representations or the infinite dimensional ones with generic $q$.

The Liouville action to start with is given by
\be
S_L(\Phi)=\frac{1}{2\pi\gamma^2}\int d^2z \sqrt{\hat{g}}
\left\{\half\hat{g}^{ab}  \partial_a \Phi \partial_b \Phi
+ \Lambda e^{\Phi(z,\zb)} \right\}
+\frac{Q_0}{2\pi\gamma}\int d^2z \sqrt{\hat{g}} R(\hat{g})\Phi(z,\zb)
\label{eq:action}
\ee
where we have chosen the metric on a Riemann surface $\Sigma$
as $ds^2=e^{\Phi(z,\zb)} \hat{g}_{z\zb} dz d\zb$ and
$\gamma$ is a coupling constant and $\Lambda$ is the cosmological
constant.
The last term in (\ref{eq:action}) is added in order for the theory
to be conformal.
Indeed, upon the Liouville equation of motion
\be
-\partial_z\partial_{\zb}\Phi(z,\zb)=-\Lambda e^{\Phi(z,\zb)},
\label{eq:eqmotion}
\ee
the energy-momentum tensor defined by
$T^L_{ab}:= 2\pi(\delta S_L/ \delta \hat{g}_{ab})$ is traceless,
$T^L_{z\zb}=0$, and satisfies the conservation law,
$\partial_{\zb}T^L_{zz}=0$.
In the derivation of the equation of motion, we have made a choice of
the background metric as $\hat{g}_{ab}=\delta_{z\zb}$.

The classical theory of the Liouville theory,
where $Q_0=Q_{cl}=1/\gamma$, yields uniformization of
a Riemann surface $\Sigma$ to the upper-half-plane
(conformally equivalent to the Poincar\'e disk).
The equation of motion (\ref{eq:eqmotion})
is solved and the solution is given by means of
arbitrary holomorphic and antiholomorphic functions $u(z), v(\zb)$
as
\be
e^{\Phi(z,\zb)}=\frac{\partial_z u(z) \partial_{\zb} v(\zb)}
{(1-u(z)v(\zb))^2},  \label{eq:solution}
\ee
which, denoting $w=u(z)$ and $\bar{w}=v(\zb)$, gives the Poincar\'e metric
\be
ds^2=\frac{dw\wedge d\bar{w}}{(1-w\bar{w})^2}
\ee
on the Poincar\'e disk $D=\{w\in{\Bbb C}\vert\, \vert w\vert<1\}$.
Thus the function $u$ and $v$ define the inverses of an uniformization map
$\jmath: D\,\rightarrow\,\Sigma$, $\jmath^{-1}(z)=w,
\overline{\jmath^{-1}}(\bar{z})=\bar{w}$.
When we quantize this system, the functions $u, v$ and, therefore,
the coordinates $w, \bar{w}$ on $D$ become operators.
In this fact we can glance at the quantum fluctuation of the metric of
the manifold, and this is the reason why we regard the quantum Liouville
theory as a theory of quantum $2D$ gravity or quantum geometry of a surface.

To study an algebraic structure of the classical Liouville theory,
let us take the $h$-th ($h\in {\Bbb Z}/2$) power of the metric
(\ref{eq:solution}).
One immediately finds there are two entirely different
cases according to whether (I) $h>0$ or (II) $h\leq 0$.
It lies in this fact that there are two kinds of quantum Liouville
gravity, \ie, of strong coupling and weak coupling regimes.
In order to see this more explicitly, we calculate the $h$-th power,
\begin{eqnarray}
({\rm I}) \quad &e^{-j\Phi(z,\zb)}& = \left(\frac{1-u(z)v(\zb)}
{\sqrt{\partial_z u(z) \partial_{\zb}v(\zb)}}\right)^{2j}=
\sum_{m=-j}^j N^j_m \psi^j_m(z)\psi^{j,m}(\zb)  \label{eq:I} \\
({\rm II}) \quad &e^{h\Phi(z,\zb)}& =
\left(\frac{\sqrt{\partial_z u(z) \partial_{\zb}v(\zb)}}{1-u(z)v(\zb)}
\right)^{2h}= \sum_{r=0}^\infty N^h_r \lambda^h_r(z)\lambda^{h,r}(\zb),
\label{eq:II}
\end{eqnarray}
where we used $h=-j$ in (\ref{eq:I}) and $N^j_m (N^h_r)$ are normalization
constants.
In the latter discussions, we are not interested in the explicit forms of
the functions $\psi$ and $\lambda$, but the crucial fact is that
$\psi^j_m(z)$ and $\lambda^h_r(z)$ form, respectively, finite and infinite
dimensional representations of $\slc$.
One can further show that the chiral sector $\psi^j_m(z)$ and
$\lambda^h_r(z)$ satisfy the Poisson-Lie relations\cite{GN,FT}.

With the above results of the classical Liouville theory in mind,
let us turn to the quantum version of the theory.
The quantization approach which many authors have
followed is the canonical quantization where equal-time commutation
relation between the Liouville field $\Phi(z,\zb)$ and its canonical
conjugate is imposed.
In the following, however, we quantize the Liouville theory
according to the well-known fact that the quantization of the Poisson-Lie
algebra ${\Got g}$ yields the quantum universal envelopping
algebra $U_q{\Got g}$.
In fact, it has been shown that the quantum Liouville theory has
the symmetry $\qslc$.
We then assume that the quantum version of the $h$-th power of
the Liouville metric, which is called the vertex operator with charge $h$,
also splits into chiral and antichiral sectors
as in (\ref{eq:I}) and (\ref{eq:II}).
The chiral sectors, denoted as $\Psi^j_m(z)$ and $\Lambda^h_r(z)$,
are now the finite and infinite dimensional representations
of $\qslc$, respectively.
The deformation parameter $q$ is given in terms of the coupling constant by
$q=\exp\,i\pi\gamma^2/2$.

The gravity in the case (I) is called the weak coupling Liouville gravity.
This case has been studied in \cite{Gerv,ST}
from the viewpoints of quantum groups.
They have shown that $\Psi^j_m(z),\, m=-j, \cdots, j$
form a finite dimensional
representation of $\qslc$ with $q=\exp\,i \pi  \gamma^2/2$ and
satisfy the braiding-commutation relations which reduce to
the Poisson-Lie relations of $\slc$ in the classical limit,
$\gamma\rightarrow0$.
In this regime, the central charge of the gravity is given by
\be
c_L=1+12Q_0^2   \label{eq:central}
\ee
and the relation between $Q_0$ and the coupling constant $\gamma$ is
\be
Q_0=\frac{1}{\gamma} +\frac{\gamma}{2}, \label{eq:QW}
\ee
which affirms that the conformal dimension of the metric $e^{\Phi(z,\zb)}$
is just $(1,1)$\cite{DK}.
The relation (\ref{eq:QW}) was also obtained in \cite{CT}
upon the assumptions that
the energy-momentum tensor should satisfy the Virasoro algebra and
$T_{zz}$ commute with $T_{\zb\zb}$.
Inserting (\ref{eq:QW}) into the central charge (\ref{eq:central}),
we get $c_L=13+6(\alpha+\alpha^{-1})$ with $\alpha=\gamma^2/2$ and
find the well-known restriction $c_L\geq25\,(\leq1)$ for real (imaginary)
$\gamma$.
If we require the central charge to be $1<c_L<25$,
the coupling constant $\gamma$ becomes a complex number.
This is the origin of the difficulty in constructing the
consistent strong coupling Liouville gravity.
The only successful work at present is the work developed by Gervais and
his collaborators\cite{Ge}, where the infinite dimensional representations of
$\qslc$ have been utilized but the deformation parameter $q$ is {\em not}
a root of unity.
It has been proved that consistent theories can be built when the central
charges take the special values, $c_L=7, 13$ or $19$.

On the contrary to the canonical quantization,
Takhtajan has recently performed path-integral quantization of the
Liouville gravity in \cite{Ta} along the Polyakov's original formulation of
quantum Liouville theory\cite{Po1}.
Upon a perturbation expansion, it is proved that the central charge of
the theory is just (\ref{eq:central}) with the classical value
$Q_0=Q_{cl}=1/\gamma$, and so $c_L$ is always greater than 1 including
strong coupling region.
It is also shown that the dimension of the metric is $(1,1)$ as desired.
Supported by the illuminating results,
let us investigate the strong coupling Liouville gravity via $\DD$, \ie,
the infinite dimensional representations with $q$
being a root of unity.

Let us go back to the case (\ref{eq:II}).
The quantum version of (\ref{eq:II}) is
\be
e^{h\Phi(z,\zb)}=\sum_{r=0}^\infty {\cal N}_q^{h,r} \Lambda^h_r(z)
\Lambda^{h,r}(\zb). \label{eq:strongvertex}
\ee
The fields $\Lambda^h_r(z), r=0, 1, \cdots$ form an infinite dimensional
lowest weight representation $V_h$ of $\qslc$ with $q=e^{\pi i \gamma^2/2}$.
In the following, let $q$ be a $p$-th root of unity, $q=e^{\pi i \qvalue}$,
\ie,
\be
\frac{\gamma^2}{2}=\qvalue, \label{eq:gvalue}
\ee
with $p, p'\in {\Bbb N}$, coprime with each other.
It is helpful to summarize here the essential features of the representation
with such a $q$ (see Refs.\cite{MS} for the detailed discussions).

A characteristic feature of the representations is that,
upon requiring the representations to be well-defined in the sense that
the norm of each vector in the lowest weight module is finite,
they are parameterized by two integers, say $\mu$ and $\nu$.
Thus the representations are discrete series and
the lowest weights are expressed in terms of $\mu, \nu$ as
\be
h_{\mu\nu}=\half \left(\frac{p}{p'}\nu-\mu+1\right)  \label{eq:lowest}.
\ee
We denote by $V_{\mu\nu}=V_{h_{\mu\nu}}$ the lowest weight module
on the lowest weight vector whose weight is $h_{\mu\nu}$
and by $\Lambda^{\mu\nu}_l$ the $l$-th vector in $V_{\mu\nu}$
where $l=0, 1, \cdots$.
One easily sees that the module $V_{\mu\nu}$ is not irreducible by itself.
Indeed, one finds infinite series of submodules in $V_{\mu\nu}$.
This is the essential difference from the infinite dimensional
representations with generic $q$.
In order to obtain the irreducible module $V^{irr}_{\mu\nu}$,
all submodules should be subtracted  correctly.
In Ref.\cite{MS} this task was done and the irreducible module
$V^{irr}_{\mu\nu}$ was obtained.
The remarkable result is that
there is an isomorphism $\rho: V^{irr}_{\mu\nu} \rightarrow V_{\zeta}^{cl}
\bigotimes\mho_j$ such that
\be
\rho\,:\, \Lambda^{\mu\nu}_{kp+r} \mapsto \lambda^{\zeta}_k
\bigotimes \Psi^j_m,  \label{eq:iso}
\ee
where $\zeta=\nu/2p'$, $j=(\mu-1)/2$ and $j+m=r$.
Therefore the irreducible lowest weight module $V^{irr}_{\mu\nu}$ is
isomorphic to the tensor product of the infinite dimensional lowest weight
module $V_{\zeta}^{cl}$ of the classical $\slc$ and
the finite dimensional highest weight module $\mho_j$  of $\qslc$,
\be
V^{irr}_{\mu\nu} \cong V_{\zeta}^{cl}\bigotimes\mho_j.  \label{eq:result}
\ee
We will make full use of this result in the following investigations of
the strong coupling Liouville gravity.

In order to apply the above results to the strong coupling Liouville gravity,
we first introduce correlation functions
defined by,
\be
Z^S[{\Got m}:\mu_i,\nu_i] =
\int[d\Phi]e^{-S_L(\Phi)}
\prod_{i=1}^N e^{h_{\mu_i \nu_i}\Phi(z_i,\zb_i)},  \label{eq:defcorr}
\ee
where ${\Got m}=(m,\overline{m})$ is the moduli parameter of the surface.
In order to obtain physical amplitude, we have to integrate $Z^S[{\Got m}]$
over the moduli space ${\cal M}_{g,N}$ with Weil-Peterson measure $d(WP)$.
Putting together (\ref{eq:iso}) and (\ref{eq:strongvertex})
with the replacement $h\rightarrow h_{\mu\nu}$, we find
\begin{eqnarray}
e^{h_{\mu\nu}\Phi(z,\zb)}&\cong&
e^{\zeta\cp(z,\zb)}\bigotimes e^{-j\vp(z,\zb)} \nonumber\\
&{}&= \sum_{k=0}^\infty N^\zeta_k \lambda^\zeta_k(z)
\lambda^{\zeta,k}(\zb)\bigotimes\sum_{m--j}^j {\cal N}^j_m
\Psi^j_m(z)\Psi^{j,m}(\zb),
\label{eq:separation}
\end{eqnarray}
where $\cp(z,\zb)$ and $\vp(z,\zb)$ are, respectively,
the classical and quantum Liouville fields.
This result suggests that the strong coupling Liouville gravity based on $\DD$
is decomposed into the classical sector and the quantum sector which
is associated with the finite dimensional representations of $\qslc$.

We now conjecture that the action also separates as
\be
S_L(\Phi)\,\mapsto\, S^{cl}(\cp) + \kappa^2 S^q(\vp)
\label{eq:saction}
\ee
with
\begin{eqnarray}
S^{cl}(\cp) &=& \frac{1}{2\pi\beta^2}\int d^2z \left(
\partial_z\cp\partial_{\zb}\cp + \Lambda_1 e^{\cp(z,\zb)}\right),
\label{eq:caction}\\
S^q(\vp) &=& \frac{1}{2\pi\gamma^2}\int d^2z \left(
\partial_z\vp\partial_{\zb}\vp + Q\gamma
R_{\hat{g}}\vp(z,\zb) + \Lambda_2 e^{\vp(z,\zb)}\right).
\label{eq:waction}
\end{eqnarray}
New coupling constant $\beta$ has been introduced for the classical sector
because, at the moment, we have no idea how it should be.
On the contrary, since, as described in (\ref{eq:result}),
the finite dimensional representation $\mho_j$
has the same deformation parameter $q$ as $V^{irr}_{\mu\nu}$,
the coupling constant for the quantum sector should be
the same as the original one, $\gamma$.
The relative constant $\kappa^2$ between the classical and quantum sectors
has been introduced and can be renormalized into
the coupling constant $\gamma$.
With (\ref{eq:separation}) and (\ref{eq:saction}),
the correlation function (\ref{eq:defcorr}) will be factorized into
classical correlation function $Z^{cl}[{\Got m}:\zeta_i]$ and
quantum one $Z^q[{\Got m}:j_i]$,
\be
Z^S[{\Got m}:\mu_i,\nu_i] \cong Z^{cl}[{\Got m}:\zeta_i]Z^q[{\Got m}:j_i].
\label{eq:correlation}
\ee

Let us discuss the result in more detail.
We first look at the classical sector briefly.
The path-integral in this sector is trivial and yields
\be
Z^{cl}[{\Got m}:\zeta_i]=e^{-\frac{c}{24\pi}\overline{S}_N^{cl}(\cp)},
\label{eq:classical}
\ee
where $c=12/\beta^2$ is the central charge of the classical Liouville gravity,
and $\overline{S}_N^{cl}(\vp)$
is the regularized action obtained by subtracting the singularities near the
points where vertices are inserted. For example, in the case when
the topology is the sphere and all vertices are punctures,
\ie, $\zeta_i=1/\beta^2$ for ${}^\forall i$, it is \cite{ZT}
\be
\overline{S}_N^{cl}(\cp) =\lim_{\epsilon\rightarrow 0}
\int_{\Sigma_\epsilon} d^2z\left( \partial_z\cp\partial_{\zb}\cp
+\Lambda_1e^{\cp(z,\zb)}
+2\pi N\log\,\epsilon+4\pi(N-2)\log\vert\log\,\epsilon\vert\right),
\label{eq:cNaction}
\ee
where $\Sigma_\epsilon=\Sigma\setminus \cup_{i=1}^{N-1}
\{\vert z-z_i\vert <\epsilon\}\cup\{\vert z\vert >1/\epsilon\}$.
We next proceed to the quantum sector.
The action $S^q(\vp)$ is just the Liouville action in the conformal gauge
$ds^2=e^{\vp(z,\zb)}dz d\zb$.
The coupling constant for this action is renormalized to
$\tilde\gamma=\gamma/\kappa$ and, therefore, the central charge
coming from this sector is
\be
c_q=1+12\tilde Q^2  \label{eq:weakcenter}
\ee
where $\tilde Q=\kappa Q$.
Note that the correlation function $Z^q[{\Got m},j_i]$
admits holomorphic factorization and can be written as
\be
Z^q[{\Got m}:j_i] =\sum_{I,J}N^{I,J}\overline{\Psi}_I[\overline{m}:j_i]
\Psi_J[m:j_i] \label{eq:holomorphic}.
\ee
Here $\Psi_I[m:j_i]$ is the Virasoro block satisfying the conformal Ward
identity obtained by Polyakov\cite{Po2} and will be regarded as a holomorphic
section of a line bundle over the moduli space ${\cal M}_{g,N}$
and $N^{IJ}$ is some constant matrix.

Let us next give a geometrical interpretation of
the strong coupling Liouville gravity.
Putting (\ref{eq:classical}) and (\ref{eq:holomorphic}) together
and integrating $Z_L[{\Got m}:\mu_i,\nu_i]$ over the moduli space
with Weil-Peterson measure, the amplitude is written as
\be
{\cal A}(\zeta_i,j_i) = \sum_{I,J}N^{I,J}
\int_{{\cal M}_{g,N}} d(WP) e^{-\frac{c}{24\pi}\overline{S}^{cl}}
\overline{\Psi}_I[\overline{m};j_i]\,\Psi_J[m:j_i].
\ee
In order to observe this result from the viewpoints of
the geometric quantization,
we should notice the fact that the Liouville action evaluated on the classical
solution is the K\"ahlar potential of the Weil-Peterson metric,
\ie, $\overline{\partial}\partial \overline{S}^{cl}_N=i\omega_{WP}/2$.
This remarkable relation was obtained in Ref.\cite{Zo} for the
topology of the $N$-punctured sphere.
Upon comparing our result with the general concept of the geometric
quantization, the holomorphic part $\Psi_I[m]$ of the
quantum sector is regarded as a holomorphic section of a line bundle
${\cal L}_c$ over the moduli space ${\cal M}_{g,N}$ with curvature given by
the K\"ahler form   $\frac{c}{48\pi}\omega_{WP}$.
On the other hand, the classical correlation function corresponds to
a Hermitian measure factor defining an inner product $\langle\,|\,\rangle_c$
in the Hilbert space which is the space of sections on ${\cal L}_c$.
Hence, at least for the topology of the $N$-punctured sphere,
the amplitude can be written as
${\cal A}=\sum N^{I,J}\langle \Psi_I |\Psi_J\rangle_c$ up to a constant.

Finally it is interesting to give a comment on the central charge.
The central charge of the original strong coupling Liouville gravity
$S_L(\Phi)$ is given by $c_L=1+12/\gamma^2$ as calculated
by Takhtajan\cite{Ta}.
On the other hand, from the right hand side of eq.(\ref{eq:correlation}),
the total central charge is the sum of the classical one, $12/\beta^2$,
and (\ref{eq:weakcenter}).
We then get the following relation among $\gamma$, $\beta$ and $\tilde Q$,
\be
\frac{1}{\gamma^2} = \frac{1}{\beta^2} + \tilde{Q}^2. \label{eq:relation}
\ee
Here we assume that $\gamma$ and $\beta$ are real.
There are three phases for the quantum sector according to
the values of the coupling constants $\gamma$ and $\beta$;
(i) $\gamma^2 > \beta^2$, (ii) $\gamma^2 < \beta^2$ and the boundary
between the two, namely,  (iii)  $\gamma^2 = \beta^2$.
We set $c_L=26-D$, \ie, $\gamma^{-1}=\sqrt{(25-D)/12}$ in order that
the central charge from the total physical system, that is, the strong
Liouville mode $\Phi(z,\zb)$ plus string sector $X^\mu(z,\zb)$ embedded in
$D$-dimensional target space, cancels against that from ghost system.
Before going to the discussions of each phases, it is convenient to rewrite
the action of the quantum sector as
\be
\kappa^2 S^w(\vp)\rightarrow
\frac{1}{2\pi}\int d^2z \left(\partial \tilde\vp
\overline{\partial}\tilde\vp+\tilde Q R_{\hat g}\tilde\vp
+\Lambda_2 e^{\tilde\vp(z,\zb)}\right),
\ee
where we have used normalized field $\tilde\vp=\vp/\tilde{\gamma}$.
In the following, we will use the notations $c=12/\beta^2$ and $D'=D+c$.
Let us see these phases separately.

\vspace{0.2cm}
\noindent
(i) In this case, $\tilde Q$ is an imaginary number.
This case corresponds to $D'>25$.
Supposing $\Lambda_2=0$ and denoting $\tilde Q=i\alpha_0$,
the quantum sector is nothing but the
minimal CFT with the background charge $\alpha_0=\sqrt{(D'-25)/12}$
and the central charge being $c_q=1-12\alpha_0^2$.
The screening charges are given by
$\alpha_\pm= (\sqrt{D'-25} \mp \sqrt{D'-1})/\sqrt{12}$.
The vertex operator $e^{-j\tilde{\gamma}\tilde{\phi}} =
e^{-ij\alpha_+\tilde{\phi}}$, which is the highest weight state
in a finite dimensional representation
of $\qslc\otimes\overline{\qslc}$, is the primary field
of the type $(2j+1, 1)\otimes\overline{(2j+1)}$.
The section $\Psi[m,j_i]$ is the conformal block of the Virasoro
minimal series.

\vspace{0.2cm}
\noindent
(ii) In this case, $\tilde Q$ is real.
This case corresponds to $D'<25$ and the quantum sector may be regarded as the
weak coupling gravity.

\vspace{0.2cm}
\noindent
(iii) Since in this case $\tilde Q=0$,
the contribution 1 in the central charge $c_L$ comes
just from the quantum sector.
This reminds us of Takhtajan's result \cite{Ta}, where he showed upon a
perturbation expansion that $12/\gamma^2=25-D$ and 1 in $c_L$ came from
the tree diagram and quantum corrections, \ie, loop diagrams, respectively.
In this case, the field $\tilde{\vp}(z,\zb)$ can be regarded as the
$(D+1)$-th component of the string coordinates, namely, $\tilde\vp=X^{D+1}$.
Remember that, in the ordinary weak coupling gravity when $D=25$,
the Liouville field can be regarded as the time component of string
in the 26-dimensional Minkowski space.
In our case, the same situation will happen for a different dimension $D$
by choosing the classical central charge $c$ appropriately.

\vspace{0.3cm}

In summary, we gave a description of a possible strong coupling
Liouville gravity via the representations $\DD$.
Our investigations were based on the fact (\ref{eq:separation})
and the {\em conjecture} (\ref{eq:saction}).
That is to say, the strong coupling Liouville gravity associated with
$\DD$ will be a total system of the classical Liouville gravity and
the quantum sector which is associated with the finite dimensional
representations of $\qslc$.
We also discussed some properties of our gravity theory.
However we have not discussed on the anomalous dimensions of the vertex
operators at all.
In order to do this within the framework of quantum groups, we have to
examine how the braiding-commutation relations in $\DD$ are decomposed into
the classical and quantum sectors.
Once we know the dimensions, we can find the relation between $\tilde Q$ and
$\tilde \gamma$ like (\ref{eq:QW}) in the weak coupling Liouville gravity.
Further studies will appear elsewhere.

In the light of our results, let us try to give an answer to
the question: {\em What is quantum space-time? }
In quantum gravity theory, every point on the space-time manifold
fluctuates by a quantum effect.
If we can regard the fluctuation as a sort of internal space on every point of
the manifold, the quantum space-time can be recognized as an object which
consists of a classical manifold and the internal spaces on it.
As for the case (i) discussed above, the quantum fluctuations, that is,
the internal spaces correspond to the minimal matter.
In this sense, we can say that the strong coupling Liouville gravity
for the case (i) is the unified theory of gravity and matter.
Recently, another unified picture of gravity and matter has been proposed
\cite{Cham} in the framework of the Connes' noncommutative
differential geometry\cite{Co}.
There, the total system is built on the space $M_4\otimes Z_2$
where $M_4$ and $Z_2$ are the 4-dimensional classical Minkowski space and
the discrete ``two-point'' space, respectively.
It must be interesting and helpful for our understanding of the
quantum space-time to search for the relationship between
the formulation and our theory.

\vspace{.5cm}
\noindent
The author would like to thank Dr. H. W. Braden
and Dr. T. Matsuzaki for valuable discussions and collaborations.

\newpage


\end{document}